\documentclass[aps,preprint,floats,nofootinbib]{revtex4}
\usepackage{amsmath}
\usepackage{amssymb}
\usepackage{latexsym}
\usepackage{epsf}
\usepackage{graphicx}

\newlength{\defbaselineskip}
\newcommand{\setlinespacing}[1]%
           {\setlength{\baselineskip}{#1 \defbaselineskip}}

\begin{document}

\hfill$\vcenter{\hbox{\bf hep-ph/0506091}}$

\vskip 0.5cm

\title {Lorentz and CPT Invariance Violation In High-Energy Neutrinos}
\author{Dan Hooper$^1$, Dean Morgan$^2$ and Elizabeth Winstanley$^2$}
\address{$^1$ University of Oxford, Oxford, United Kingdom;\\
$^2$ University of Sheffield, Sheffield, United Kingdom}
\date{\today}

\bigskip

\begin{abstract}

High-energy neutrino astronomy will be capable of observing particles at both extremely high energies and over extremely long baselines. These features make such experiments highly sensitive to the effects of CPT and Lorentz violation. In this article, we review the theoretical foundation and motivation for CPT and Lorentz violating effects, and then go on to discuss the related phenomenology within the neutrino sector. We describe several signatures which might be used to identify the presence of CPT or Lorentz violation in next generation neutrino telescopes and cosmic ray experiments. In many cases, high-energy neutrino experiments can test for CPT and Lorentz violation effects with much greater precision than other techniques.

\end{abstract}

\maketitle

\section{Introduction}

The invariance of the product of Charge conjugation, Parity and
Time reversal (CPT) is one of the most fundamental symmetries
found in physics \cite{PCT}. The invariance of CPT is also
intimately related with Lorentz invariance, another symmetry of
deeply embedded within our current picture of the physical world.

It is not completely clear that these symmetries will
remain perfectly unbroken under all conditions, however. In
particular, many efforts to describe the force of gravity within
the context of a quantum theory imply the breaking of CPT and
Lorentz symmetries. These effects are so minuscule, however, that
any indication of them will likely require observations either at extremely high energies, or over incredibly long baselines. High energy neutrino astronomy provides precisely such a laboratory.

Most varieties of particles cannot travel undisturbed over cosmological distances. Particles which scatter off
of background radiation or other targets are of limited
use to very long baseline experiments, particularly at very
high energies. High-energy neutrinos, on the other hand, can
travel thousands of mega-parsecs without undergoing any energy
loses beyond those from Hubble expansion.

Experimentally speaking, the field of high-energy neutrino
astronomy is developing very rapidly. Current technology, such as the
AMANDA II \cite{amanda2} and Rice \cite{rice} telescopes at the
South Pole, have been operating successfully for several years.
The construction of next generation optical Cerenkov neutrino
experiments such as IceCube \cite{icecube} and Antares
\cite{antares} is also going ahead. Additionally, ultra-high cosmic ray experiments such as Auger \cite{auger}, EUSO \cite{euso} and OWL \cite{owl} will be very sensitive to the highest energy range of the cosmic neutrino spectrum. The prospects for other
technologies which incorporate radio \cite{anita} and acoustic
\cite{acoustic} detectors appear to be very promising as well. With
currently existing experiments collecting data and new experiments being developed and constructed, a new window into the Universe is beginning to open.

Over very long distances, the effects of CPT and Lorentz violation
can result in modifications to the standard neutrino oscillation
phenomenology. By using high-energy neutrino telescopes to measure
the ratios of neutrinos flavors coming from distant sources, the
possible effects of CPT and Lorentz violation can be constrained
far beyond the levels which are currently experimentally accessible.

In this article, we discuss how high-energy cosmic neutrinos can
be used to test for the violation of CPT and Lorentz symmetries.
In section~\ref{sec:Nuosc}, we review the present status of
standard neutrino oscillation phenomenology and discuss the
violation of Lorentz and CPT symmetries in
section~\ref{sec:VioLVCPT}. In section~\ref{pheno} we discuss
these effects in the context of neutrino oscillation
phenomenology, and describe potential sources and experimental
prospects for the observations of these effects in
section~\ref{mysec}. We summarize our conclusions in section~\ref{conc}.

\section{Standard Neutrino Oscillations: Present Status} \label{sec:Nuosc}
\subsection{Formalism}\label{Nuoscform}
\subsubsection{Vacuum oscillations}
Normally, neutrinos are identified by their flavor ($e$, $\mu$,
$\tau$) rather than their mass. Neutrinos which have
definite flavor need not be in states of definite mass, however. If this is
the case, we consider the flavor states, $|\nu_{\alpha}\rangle$,
to be a linear combination of mass states, $|\nu_{i}\rangle$:
\begin{equation}
\label{super} |\nu_{\alpha}\rangle=\sum_{i}U_{\alpha
i}|\nu_{i}\rangle,
\end{equation}
where $U_{\alpha i}$ are the components of the unitarity leptonic
mixing matrix. In the Schr\"{o}dinger representation, the time
evolution of the mass eigenstates has the form
\begin{equation}
\label{eqn:schro} i\frac{d}{d
\tau}|\nu_{i}(\tau)\rangle=m_{i}|\nu_{i}(\tau)\rangle,
\end{equation}
where $\tau$ is the time in the mass frame and $m_{i}$ the mass
eigenvalues. Using Eq.~(\ref{eqn:schro}), the probability of
oscillation from $\nu_{\alpha}$ to $\nu_{\beta}$ can be calculated
to be
\begin{eqnarray}
\label{prob:stan}
\nonumber  P(\nu_{\alpha} \rightarrow \nu_{\beta}) &=&
   \delta_{\alpha\beta} -4\sum_{j>i}\Re(U_{\alpha j}^{*}U_{\beta j}U_{\alpha i}U_{\beta i}^{*})\sin^{2}[\Delta
   m_{ji}^{2}(L/4E)] \\
   &  & \quad \:\:\, +2\sum_{j>i}\Im(U_{\alpha j}^{*}U_{\beta j}U_{\alpha i}U_{\beta i}^{*})\sin[\Delta
   m_{ji}^{2}(L/2E)],
\end{eqnarray}
where $\Delta m_{ji}^{2}\equiv m_{j}^{2}-m_{i}^{2}$, $E$ is the
energy of the neutrino and $L$ is the path length. We have assumed
the neutrino to be relativistic. If we consider three flavors of
Majorana neutrinos, then the $3\times 3$ mixing matrix is
given by $U=V \cdot M$, where
\begin{equation}
\nonumber
\label{eqn:3x3mix} V =\left(%
\begin{array}{ccc}
  c_{12}c_{13} & s_{12}c_{13} & s_{13}e^{-i\delta_{CP}} \\
  -s_{12}c_{23}-c_{12}s_{23}s_{13}e^{i\delta_{CP}} & c_{12}c_{23}-s_{12}s_{23}s_{13}e^{i\delta_{CP}} & s_{23}c_{13} \\
  s_{12}s_{23}-c_{12}c_{23}s_{13}e^{i\delta_{CP}} & -c_{12}s_{23}-s_{12}c_{23}s_{13}e^{i\delta_{CP}} & c_{23}c_{13} \\
\end{array}%
\right),
\end{equation}
\begin{equation}
M=\left(%
\begin{array}{ccc}
  e^{i\phi_{1}} & 0 & 0\\
  0 & e^{i\phi_{2}} & 0\\
  0 & 0 & 1\\
\end{array}%
\right).
\end{equation}
Here, $c_{ij}$ and $s_{ij}$ represent the cosine and sine of the
mixing angle $\theta_{ij}$ respectively, $\delta_{CP}$ is a CP
violating phase and the phases in $M$ are the Majorana phases. For
Dirac neutrinos, the situation is similar but the Majorana phases
may be absorbed into the phases of the mass eigenstates. In
practice, since the mixing angle $\theta_{13}$ has been found to
be small, systems involving just two neutrinos are often considered. Then,
replacing $c$ and $\hbar$, the vacuum oscillation probability
reduces to
\begin{equation}
\label{eqn:standprob2} P(\nu_{\alpha} \rightarrow \nu_{\beta}) =
\sin^{2}2\theta\sin^{2}\left[1.27\Delta m^{2}\frac{L}{E}\right],
\end{equation}
where $\Delta m^{2}$ is measured in eV$^{2}$, $L$ is measured in
kilometers and $E$ is measured in GeV. By convention, $\theta_{12}$
and $\Delta m^{2}_{21}$ are solar oscillation parameters
describing $\nu_{e}\rightarrow\nu_{\mu,\tau}$ whilst $\theta_{23}$
and $\Delta m^{2}_{32}$ are atmospheric neutrino oscillation
parameters, describing $\nu_{\mu}\rightarrow\nu_{\tau}$.
\subsubsection{Oscillations in matter}
The situation becomes somewhat more complicated if neutrinos are
passing through dense matter, for example in the Sun. In this
case, the Hamiltonian in the mass basis is no longer diagonal, resulting in an energy dependence in the mixing angles.
For simplicity, we will consider a two neutrino system.
Incorporating matter effects, the Hamiltonian in the mass basis is
\cite{zuber}
\begin{equation}
\label{eqn:hamMEMB} H^{i}_{m}=\frac{1}{2E}\left(%
\begin{array}{cc}
  m_{1}^{2}+A\cos^{2}\theta & A\sin\theta\cos\theta \\
  A\sin\theta\cos\theta & m_{2}^{2}+A\sin^{2}\theta \\
\end{array}%
\right),
\end{equation}
where $A$, embodying the matter effects, is
$A=2\sqrt{2}EG_{F}N_{e}$ with $G_{F}$ the Fermi constant and
$N_{e}$ the electron number density. The oscillation probability
therefore becomes
\begin{equation}
\label{eqn:MSWprob} P(\nu_{e} \rightarrow \nu_{\beta}) =
\sin^{2}2\theta_{m}\sin^{2}\left[\Omega_M L\right].
\end{equation}
where
\begin{equation}
\label{eqn:MSWcomp} \sin2\theta_{m}=\frac{\Delta
m^{2}}{4E}\frac{\sin2\theta}{\Omega_M},
\end{equation}
with
\begin{equation}
\label{eqn:MSWOmega} \Omega_M=\frac{\Delta
m^{2}}{4E}\sqrt{\left[\left(\frac{A}{\Delta
m^{2}}-\cos2\theta\right)^{2}+\sin^{2}2\theta\right].}
\end{equation}
For a comprehensive overview of neutrino physics, see for example
Ref.~\cite{zuber}.

\subsection{Where we stand - the status of neutrino oscillations}
\subsubsection{Solar neutrino oscillations}
The first indirect evidence for neutrino oscillations came from
those neutrinos created in the Sun. The standard solar model
\cite{bahcall} describes the complex nuclear processes occurring
in the Sun and from this, predictions of the solar neutrino flux
may be extracted. Early experiments measuring the solar
neutrino flux, such as the Chlorine \cite{chlorine} and Gallium
\cite{gallium}, were sensitive to electron neutrinos only and
reported a neutrino flux significantly less than those predicted
by the standard solar model. More recently, the SNO \cite{sno}
experiment, which is sensitive to all neutrino flavors, showed
the total solar neutrino flux agrees well with the predictions of
the standard solar model but with an appreciable suppression of
the electron neutrino flux. This suppression is definitive
evidence of neutrino oscillations from $\nu_{e}\rightarrow\nu_{\mu
,\tau}$. The KamLAND experiment \cite{kamland}, which detects
electron anti-neutrinos from nuclear reactors, reported a
significant suppression in event rates thus corroborating the SNO
results. Using data from these experiments, the solar neutrino oscillation parameters, $\Delta
m^{2}_{21}=\Delta m^{2}_{\odot}$ and
$\sin^{2}\theta_{21}=\sin^{2}\theta_{\odot}$, have been measured with high precision.
These values are shown in table \ref{tab:oscparam}.
\subsubsection{Atmospheric neutrino oscillations}
Whilst the first indirect evidence for neutrino oscillations came
from solar neutrinos, the first direct evidence came from
atmospheric neutrinos. Atmospheric neutrinos are created by
interactions of cosmic rays with atmospheric atomic nuclei. Pions,
created in this interaction decay by
$\pi^{-}\rightarrow\mu^{-}+\overline{\nu}_{\mu}\rightarrow\overline{\nu}_{\mu}+\nu_{\mu}+\overline{\nu}_{e}+e^{-}$
(with an analogous decay chain for $\pi^{+}$), thus indicating
that the muon neutrino flux should be roughly twice that of
electron neutrinos. However, the Kamiokande \cite{kamio}, IMB
\cite{imb} and Soudan \cite{soudan} experiments reported a
significant deficit in the expected $\nu_{\mu}:\nu_{e}$ ratio
\cite{kamiores,imbres,soudanres} which suggested oscillations from
$\nu_{\mu}$ to $\nu_{\tau}$. These results were somewhat inconclusive, however. In 1998, this
changed when the Super-Kamiokande collaboration showed the muon
neutrino flux had a zenith angle dependence, which implied a
dependence upon path length \cite{newsuperk}. This was the first direct evidence
that neutrinos oscillate from one flavor to another. A summary of
the present values of the atmospheric neutrino oscillation
parameters is given in table \ref{tab:oscparam}.
\subsubsection{Three neutrino oscillations}
The standard analysis of solar and atmospheric neutrino flux is
done within the two neutrino approximation. However, in order to
place values on $\Delta m^{2}_{31}$ and $\sin^{2}\theta_{31}$, the
data must be analyzed taking all three neutrino flavors into
account. Table \ref{tab:oscparam} again shows the present status.
\subsubsection{The LSND result}
It would seem from the discussion above that, short of higher precision measurements of the oscillation parameters being performed, the phenomenology of neutrino
oscillations is well understood. However, this is not
entirely the case. The results of the LSND experiment \cite{lsnd}, which produces a beam of $\nu_{e}$, $\nu_{\mu}$ and
$\overline{\nu}_{\mu}$ and then searches for the appearance of
$\overline{\nu}_{e}$ that have oscillated from $\overline{\nu}_{\mu}$, cannot be reconciled with the solar and atmospheric neutrino data. The LSND results imply a mass difference which lies in the range of
$0.2<\Delta m^{2}_{LSND}<10$ $\rm{eV}^{2}$ \cite{lsndres}. If this
result were corroborated by the miniBoone experiment
\cite{miniboone}, then it would provide indications of new
physics. In particular, in order to combine in a compatible way the LSND result with the atmospheric and solar oscillation data, one would have to
invoke oscillations into sterile neutrinos or break CPT invariance. This latter possibility is discussed below.
\begin{table}[t!]
\begin{center}
\begin{tabular}{|lccl|}
\hline
Parameter & Value  & 2/3 neutrino analysis& data\\
 \hline
 $\Delta m_{\odot}^{2}$ & $8.3\times10^{-5}$~eV$^{2}$& 2& solar + KL(766.3 TY) \\
 $\Delta m_{atm}^{2}$  & $2.3\times10^{-3}$~eV$^{2}$ & 2 & SK + K2K\\
  $ \sin^{2}\theta_{\odot} $ & 0.27 &  2 & solar + KL(766.3 TY)\\
  $\sin^{2} \theta_{atm} $  & $0.50$ &  2 & SK + K2K\\
$\sin^{2}\theta_{13}$  & $<0.05$&  3 & solar+KL(766.3 TY)\\
 & & & + CHOOZ + atm\\
\hline
\end{tabular}
\end{center}
\caption{Current status of neutrino oscillation parameters found
from the Super-Kamiokande (SK), K2K, KamLAND, (KL) and CHOOZ solar
and atmospheric neutrino experiments. These values have been taken
from Ref.~\cite{goswami}.} \label{tab:oscparam}
\end{table}
\section{Violating Lorentz and CPT
invariance}\label{sec:VioLVCPT} The issues of CPT invariance
violation (CPTV) and Lorentz invariance violation (LV) are
intimately related. The CPT theorem is a fundamental ingredient of
quantum field theory ensuring that quantities appearing in the
theories, such as the Hamiltonian and Lagrangian density, are
invariant under the combined operations of charge conjugation (C),
parity reflection (P) and time reversal (T). The CPT theorem
holds in flat space-times provided the theory obeys
\begin{itemize}
\item locality, \item unitarity, \item Lorentz invariance.
\end{itemize}
Deviation from any one of these requirements leads to CPTV. It has
also been shown recently that CPTV leads to LV \cite{greenberg}.
\subsection{Violations of Lorentz invariance}
The breaking of Lorentz symmetry may arise as a consequence of
quantum gravity from non-trivial effects at the Planck scale, such
as the existence of a fundamental length scale. Naively, one would
expect this length to be the same in all reference frames due to
its fundamental nature. This invariance between frames is in
direct disagreement with special relativity, the Lorentz
transformation predicting length contraction. Thus, there are
three possibilities for the fate of LV. Firstly, Lorentz
invariance may hold at the Planck scale, so that the flat
space-time picture is valid. In this case, it is our naive thought
experiment which is wrong. Secondly, Lorentz symmetry may be
broken at the Planck scale suggesting a class of preferred
inertial frames. Very often, this preferred frame is assumed to be
the related to the cosmic microwave background. In this case, the
dispersion relation for energy and momentum is modified and
depends upon the quantum gravity environment. Thirdly, Lorentz
symmetry may be deformed \cite{camelia}. Again, this leads to a
modified dispersion relation but with the Lorentz transformations now
containing a second observer independent scale. For example, the
Planck length could be an independent scale in addition to the speed of
light.
\subsubsection{Lorentz invariance violation in string theory}
String theory approaches the quantum gravity problem from a
particle physics perspective. Although string theory does not quantize
space-time since it describes the background space-time entirely
classically, the issues related to this are still far from being resolved.
At this moment in time, since the background is classical, there
are no indications that Lorentz invariance is broken within string
theory. Of course, if it is found that the background space-time
needs to be quantized, then this could lead to Lorentz invariance
violating string theories. Having said that, two particular
theories which are considered to be low energy limits of string
theory, namely flat non-commutative space-times \cite{madore} and
the Standard Model Extensions (SMEs) \cite{kostelecky}, indicate the presence of LV. The non-commutative space-time approach assumes that
space-time coordinates do not commute, leading to the breaking of
Lorentz symmetry and various forms of the dispersion relation
\cite{minwalla,matusis}. The SMEs extend the Standard Model
Lagrangian to include all LV and CPTV operators which are of
dimension 4 or less, in order to be renormalizable. Again, this
phenomenological model modifies the dispersion relation.
\subsubsection{Lorentz invariance violation in loop quantum gravity}
In contrast to string theory, loop quantum gravity (also known as
canonical quantum gravity) approaches the quantum gravity problem
from a general relativity perspective. The theory is fully
background independent, as is general relativity, which leads
to the prediction of discrete space-times \cite{rovelli}.
Initially, it was thought that loop quantum gravity would preserve
Lorentz symmetry, however it is now thought that this theory
could break \cite{gambini,alfaro,thiemann} or deform
\cite{camelia2,freidel} Lorentz symmetry, thus leading to modified
dispersion relations. The issue of whether Lorentz symmetry is
preserved, broken or deformed is still unresolved in loop quantum
gravity.
\subsection{Violations of CPT}
The breaking of CPT invariance may occur independently of LV
effects from the loss of unitarity leading to quantum decoherence.
Again, this loss of unitarity could occur because of the discrete
and topologically non-trivial nature of space-time resulting in
the vacuum creation of quantum black holes with event horizons
having radii of order the Planck length. This continuous creation
and evaporation of these quantum singularities results in
space-time having a foamy nature. When particles pass by these
quantum black-holes, some of the particles' quantum numbers could
be captured by the space-time fluctuations. With the evaporation
of the black holes, the information captured would be lost to the
vacuum, inaccessible to low energy experiments. This loss of
information would imply that initially pure quantum states may
evolve into mixed quantum states; a process forbidden within
standard quantum mechanics. Since we are now dealing with mixed
quantum states, we use the density matrix formalism. Including
decoherence linearly, the time evolution of the density matrix,
$\rho$, is modified:
\begin{equation}
\label{eqn:densmatTE} \frac{d\rho}{dt}=-i[H,
\rho]+\mathcal{D}[\rho],
\end{equation}
where $H$ is the Hamiltonian of the system interacting with the
environment through the operators $D_{j},D_{j}^{\dag}$. We would
expect the the CPTV term, $\mathcal{D}[\rho]$, to take the Lindblad
form \cite{lindblad}
\begin{equation}
\label{eqn:lind}
\mathcal{D}[\rho]=\sum_{j}\left(\{\rho,D_{j}^{\dag}D_{j}\}-2D_{j}D_{j}^{\dag}\right)
\end{equation}
where $\{...\}$ represents an anti-commutator. From a physical
point of view, we require energy conservation and monotonic
increase in the von-Neumann entropy. In this case, we find we have
to specify the operators, $D$, to be self-adjoint and that they
commute with the Hamiltonian. We therefore find
\begin{equation}
\label{eqn:Dcommutator}
\mathcal{D}[\rho]=\sum_{j}[D_{j},[D_{j},\rho]].
\end{equation}
From a quantum gravity perspective, we would expect the operators,
$D$, to be proportional to the inverse of the Planck mass and thus $\mathcal{D}[\rho]\propto M_{p}^{-2}$.
\subsubsection{Quantum decoherence in string theory}
The evolution of mixed states into pure states, as described
above, results in problems with defining an $S$ matrix. Since string
theory relies on the defining of $S$ matrices, quantum
decoherence is generally not expected within string theory. However, one class of string
theories, namely non-critical string theories, may allow
decoherence. This theory may be viewed as a type of
non-equilibrium string theory with the so called critical strings
corresponding to equilibrium points within the theory (for more
details see, for example, Ref.~\cite{noncritical}). In this case, we
find an analogous expression for the time evolution of the string
matter density matrix as with Eq.~(\ref{eqn:Dcommutator})
\cite{mavromatos2}.
\subsubsection{Quantum decoherence in loop quantum gravity}
Whilst loop quantum gravity implies that space-time is discrete,
there is no a priori reason to expect quantum decoherence.
However, there have been proposals \cite{gambini2} suggesting that
the discreteness of space-time may induce decoherence having the
Lindblad form outlined in Eq.~(\ref{eqn:densmatTE}). However,
it seems that there is still much work needed in order to clarify
this.
\subsubsection{Cosmological decoherence}
In addition to quantum decoherence induced by space-time foam
effects, it may be that decoherence arises from cosmological
considerations. It is now established that the universe has
entered a period of acceleration \cite{supernova1,supernova2}
driven by some exotic dark energy. If this expansion continues,
the universe will evolve into a de-Sitter universe, expanding at
an exponential rate. This would imply the existence of a
cosmological horizon. This situation can be considered in the same
way as that of the space-time foam situation except we, as the
observers, now inhabit the space within the horizon instead of
outside. The existence of this horizon would again lead to the
inability to define $S$ matrices leading to decoherence. It has
been argued in non-critical string theory \cite{mavromatos3} that
this cosmological decoherence may be intimately linked with
quantum gravity. Considering a two level neutrino system, the
cosmological decoherence parameter, $\gamma_{cosmo}$, is related to
the cosmological constant, $\Lambda$, the weak string coupling,
$g_{s}$, the difference of the squares of the mass eigenstates,
$\Delta m^{2}$, the energy of the neutrino, $E$, and the string
mass scale, $M_{s}$:
\begin{equation}
\label{eqn:cosmodeco} \gamma_{cosmo}\sim\frac{\Lambda
g_{s}^{2}(\Delta m^{2})^{2}}{E^{2}M_{s}}.
\end{equation}
\section{Neutrino oscillation phenomenology with CPTV and LV}
\label{pheno}
\subsection{Neutrino oscillations and LV effects}
\subsubsection{Modified dispersion relations} As discussed above,
if we allow LV, then this leads to modified dispersion relations
(MDR). From the discussions above, we find it useful to
parameterize the MDR's, to leading order in the Planck energy,
$E_{p}$, as
\begin{equation}
\label{eqn:moddisp} E^{2}=p^{2}+m^{2}+\eta
p^{2}\left(\frac{E}{E_{p}}\right)^{\alpha}
\end{equation}
where $E$ is the energy of the neutrino, $p$ is its momentum, $m$
is the mass eigenstate and $\eta$ and $\alpha$ are LV parameters.
Assuming that that the parameter $\eta$ is not universal and
depends upon the mass eigenstate, we may write the two neutrino
Hamiltonian in the mass basis as
\begin{equation}
\label{eqn:2neutLVHam} H=\left(%
\begin{array}{cc}
  \frac{m_{1}^{2}}{2E}+\frac{\eta_{1}{E^{\alpha+1}}}{2} & 0 \\
  0 & \frac{m_{2}^{2}}{2E}+\frac{\eta_{2}{E^{\alpha+1}}}{2} \\
\end{array}%
\right),
\end{equation}
where we have neglected the kinetic term and identified $p$ with
$E$ since the mass eigenstates and the LV are terms are much
smaller than the momentum of the neutrinos. For simplicity, we
have absorbed the Planck energy into $\eta$. This leads to the
neutrino oscillation probability,
\begin{equation}
\label{eqn:2neutLVprob}
P[\nu_{\alpha}\rightarrow\nu_{\beta}]=\sin^{2}2\theta\sin^{2}\left[\frac{\Delta
m^{2}L}{4E}+\frac{\Delta\eta E^{n}L}{4}\right],
\end{equation}
where $n=\alpha+1$ and $\Delta \eta$ is the difference between the two values of $\eta$. If there are no LV effects, we recover the
standard neutrino oscillation probability (\ref{eqn:standprob2})
(replacing $c$ and $\hbar$). We also assumed that the LV parameter,
$\eta$, had a dependence on the mass eigenstate. If this is not the
case, then the neutrino probability remains invariant even if
Lorentz invariance is violated or deformed. Assuming that these
effects take place in atmospheric neutrino oscillations, the LV
effects become significant when
\begin{equation}
\label{eqn:2neutLVsig} 1.27\frac{\Delta m^{2}L}{E}\sim
1.27\times10^{27}\Delta\eta E^{2}L
\end{equation}
where we have set $\alpha=1$ for simplicity. We therefore find
\begin{equation}
\label{eqn:2neutLVbound} \Delta\eta\sim\frac{\Delta
m^{2}}{10^{27}E^{2}}\sim10^{-30}~eV^{-1}
\end{equation}
using the value for $\Delta m^{2}$ from table \ref{tab:oscparam}
and $E=1~\rm{GeV}$, the peak in the atmospheric neutrino flux.
\subsubsection{Neutrino oscillations and the SME}
As we outlined above, the Standard Model Extension (SME) can be
considered as a low energy phenomenological model of string theory.
The effective SME Hamiltonian describing
flavor neutrino propagation, to first order, is
\cite{kostelecky2}
\begin{equation}
\label{eqn:SMEgenHam}
H_{\alpha\beta}^{eff}=|\overrightarrow{p}|\delta_{\alpha\beta} +
\frac{1}{2|\overrightarrow{p}|}
  [\widetilde{m}^{2}+2(a_{L}^{\mu}p_{\mu}-(c_{L})^{\mu\nu}p_{\mu}p_{\nu})]_{\alpha\beta}
\end{equation}
where $\widetilde{m}$ is related to the standard neutrino mass, $\alpha$,
$\beta$ are flavor indices and $a_{L}$, $c_{L}$ violate Lorentz
invariance and CPT invariance. One of the main differences between
this model and the LV effects described in the last section is that the
Hamiltonian need not be diagonal. In Ref.~\cite{kostelecky2},
various assumptions are made in order to simplify the model. Here, instead, we will adopt a general off-diagonal formalism.
\par
In the two neutrino case, we will assume a Hamiltonian in the mass
basis of the form
\begin{equation}
\label{eqn:2neutoffdiagH} H_{eff}=\left(%
\begin{array}{cc}
  \frac{m_{1}^{2}}{2E} & a_{1}-ia_{2} \\
  a_{1}+ia_{2} & \frac{m_{2}^{2}}{2E} \\
\end{array}%
\right),
\end{equation}
where $a_{1}$ and $a_{2}$ are real, off-diagonal, LV parameters
(the $a$'s here are independent of the $a$ in Eq.~(\ref{eqn:SMEgenHam})).
In order to calculate the probability, we will use the density
matrix formalism. Writing the Hamiltonian in terms of the Pauli
matices gives
\begin{equation}
\label{eqn:2neutnondiagHab} h_{ij}=-2\left(%
\begin{array}{ccc}
  0 & -\frac{\Delta m^{2}}{4E} & -a_{2} \\
  \frac{\Delta m^{2}}{4E} & 0 & a_{1} \\
  a_{2} & -a_{1} & 0 \\
\end{array}%
\right)
\end{equation}
where we have omitted the zeroth components for simplicity as they
are all identically zero. This matrix has eigenvalues,
$\lambda_{i}$, given by $\{\pm i\Omega,~0\}$ where
$\Omega=\sqrt{\omega^{2}+a_{1}^{2}+a^{2}_{2}}$ and $\omega=\Delta
m^{2}/4E$ and the matrix in Eq.~(\ref{eqn:2neutnondiagHab})
is diagonalized by the unitary matrix
\begin{equation}
\label{eqn:2neutnondiagU} U=\frac{1}{\sqrt{2a_{1}^{2}+2a_{2}^{2}}\Omega}\left(%
\begin{array}{ccc}
  \omega a_{1}-ia_{2}\Omega & \omega a_{1}+ia_{2}\Omega & -a_{1}\sqrt{2a_{1}^{2}+a_{2}^{2}} \\
  \omega a_{2}+ia_{1}\Omega & \omega a_{2}-ia_{1}\Omega & -a_{2}\sqrt{2a_{1}^{2}+a_{2}^{2}} \\
  a_{1}^{2}+a_{2}^{2} & a_{1}^{2}+a_{2}^{2} & \omega\sqrt{2a_{1}^{2}+a_{2}^{2}} \\
\end{array}%
\right).
\end{equation}
The components of the density matrix are given by
\begin{equation}
\label{eqn:2neutnondiagrho}
\rho_{i}(L)=\sum_{j,k}U_{ij}e^{\lambda_{j}L}U_{jk}^{-1}\rho_{k}(0),
\end{equation}
where $U_{ij}$ are components of the matrix in equation
(\ref{eqn:2neutnondiagU}) and $\rho(0)$ is the density matrix
initially. Assuming we have a muon neutrino which oscillates into
a tau neutrino, the probability of oscillation is given by
\begin{equation}
\label{eqn:2neutnondiagP} P=Tr[\rho_{\mu}(L)\rho_{\tau}(0)]
\end{equation}
with
\begin{equation}
\label{eqn:2neutnondiagRi} \rho_{\mu}(0)=\left(%
\begin{array}{cc}
  \cos^{2}\theta & \sin\theta\cos\theta \\
  \sin\theta\cos\theta & \sin^{2}\theta \\
\end{array}%
\right),\quad\quad \rho_{\tau}(0)=\left(%
\begin{array}{cc}
  \sin^{2}\theta & -\sin\theta\cos\theta \\
  -\sin\theta\cos\theta & \cos^{2}\theta \\
\end{array}%
\right).
\end{equation}
Thus the probability of oscillation is
\begin{eqnarray}
\label{eqn:2neutnondiagOP}\nonumber
P[\nu_{\mu}\rightarrow\nu_{\tau}]
&=&\frac{1}{2}\left[\cos^{2}2\theta\left(1-\frac{\omega^{2}}{\Omega^{2}}-\frac{|a|^{2}}{\Omega^{2}}\cos(2\Omega
L)\right)\right.\\
\nonumber
&&\left.+\sin^{2}2\theta\left(1-\frac{a_{1}^{2}}{\Omega^{2}}-\frac{(\omega^{2}+a_{2}^{2})}{\Omega^{2}}\cos(2\Omega
L)\right)\right.\\
&&\left.-\frac{1}{2}\sin4\theta\left(-\frac{4\omega
a_{1}}{\Omega^{2}}\sin^{2}(\Omega L)\right)\right],
\end{eqnarray}
with $a=a_{1}+ia_{2}$. In an analogous way to the diagonal case,
the LV parameter, $a$, may have an explicit dependence on the
neutrino energy. To examine this energy dependence, we let
$a\rightarrow aE^{n}$ where $n$ is an extra parameter of the
theory. It is also particularly interesting to note that if we let
the quantities in Eq.~({\ref{eqn:2neutoffdiagH}) go to
\begin{eqnarray}
\label{eqn:2neutLVmatcomp} \nonumber m_{1}^{2}&\rightarrow&
m_{1}^{2}+A\cos^{2}\theta,\\
\nonumber m_{2}^{2}&\rightarrow&
m_{1}^{2}+A\sin^{2}\theta,\\
\nonumber a_{1} &\rightarrow&
\frac{A}{2E}\sin\theta\cos\theta,\\
a_{2}&=&0,
\end{eqnarray}
then we recover the Lorentz invariant matter effects situation as
described in section \ref{sec:Nuosc}.
\par
If we wish to include these off-diagonal LV effects in the case
for three neutrinos, the situation becomes very difficult as, we have three mixing angles, three mass differences and
three LV parameters. In order to examine how the LV effects
manifest themselves, we consider only a first order approximation
in the LV parameters. In the standard oscillation case, the time
evolution of the density matrix is given by
\begin{equation}
\label{eqn:3neutoffdiagTE} \frac{d\rho}{dt}=B\rho
\end{equation}
where $B$ is the matrix representing the Hamiltonian in the Pauli
basis. Perturbing the density matrix and the matrix $B$:
\begin{eqnarray}
\label{eqn:3neutoffdigpert} \nonumber \rho&\rightarrow&\rho_{0}+\delta\rho_{1}\\
B&\rightarrow&B+\delta C,
\end{eqnarray}
where the $\delta$ quantities contain the LV effects. Substituting
(\ref{eqn:3neutoffdigpert}) into (\ref{eqn:3neutoffdiagTE}) and
equating coefficients gives
\begin{equation}
\label{eqn:3neutoffdiagTEP}
\delta\dot{\rho}_{1}=B\delta\rho_{1}+\delta C\rho_{0}.
\end{equation}
Defining the vectors $\textbf{x},~\textbf{y}$ as
\begin{equation}
\label{eqn:3neutoffdiagxy}
\mbox{\boldmath$\rho_{0}$\unboldmath}=U{\bf{x}},\quad\quad
\delta\mbox{\boldmath$\rho_{1}$\unboldmath}=U{\bf{y}},
\end{equation}
where the components of the $\rho$ vectors are the components of
the density matrix and $U$ is the unitary matrix diagonalizing
$B$, we may rewrite Eq.~(\ref{eqn:3neutoffdiagTEP}) as
\begin{equation}
\label{eqn:3neutoffdiagDE}
\mbox{\boldmath$\dot{\rm{y}}$\unboldmath}-U^{-1}BU\mathbf{y}=U^{-1}\delta
CU\mathbf{x}.
\end{equation}
Since we know $U,~B,~\delta C$ and can evaluate $\mathbf{x}$, then
solving this equation will give us the perturbation to the density
matrix from which we may calculate oscillation probabilities. In
reality, this calculation still results in complicated expressions
for the probabilities. However, the expressions are greatly
simplified if we assume very long path lengths. This is entirely
reasonable since we need only consider the three neutrino system
when considering neutrinos from astrophysical sources. Using the
values of the mixing parameters from table \ref{tab:oscparam},
assuming a normal mass hierarchy and that the LV parameter, $a$,
is real, we find the oscillation probabilities to be
\begin{eqnarray}
\label{eqn:3neutoffdiagPROB} \nonumber
P[\nu_{e}\rightarrow\nu_{e}] &=&
0.564-4.39\times10^{9n+12}aE^{n+1},\\
\nonumber P[\nu_{e}\rightarrow\nu_{\mu}] &=&
0.264+1.54\times10^{9n+12}aE^{n+1},\\
\nonumber P[\nu_{e}\rightarrow\nu_{\tau}] &=&
0.180+2.93\times10^{9n+12}aE^{n+1},\\
\nonumber P[\nu_{\mu}\rightarrow\nu_{\mu}] &=&
0.365-1.30\times10^{9n+11}aE^{n+1},\\
\nonumber P[\nu_{\mu}\rightarrow\nu_{\tau}] &=&
0.367-1.16\times10^{9n+12}aE^{n+1},\\
P[\nu_{\tau}\rightarrow\nu_{\tau}] &=&
0.449-1.56\times10^{9n+12}aE^{n+1},
\end{eqnarray}
where we have written out in full the explicit dependence of the
LV parameter on the neutrino energy and replaced $c$ and $\hbar$. These results are only valid near the threshold of LV effects setting in. 
Using these probabilities, it is possible to find expressions
describing the flux of neutrinos originating in astrophysical
sources. If we assume that only electron and muon neutrinos are
created, we parameterize the initial flux as
\begin{eqnarray}
\label{eqn:3neutoffdiagIF} \nonumber \Phi_{e}&=&\varepsilon
\Phi_{tot},\\
\Phi_{\mu}&=&(1-\varepsilon)\Phi_{tot}
\end{eqnarray}
where $\varepsilon \in [0,1]$ and $\Phi_{tot}$ is the total flux.
In terms of the neutrino probabilities of Eq.~(\ref{eqn:3neutoffdiagPROB}), the neutrino flavor composition at
the detector is given by
\begin{eqnarray}
\label{eqn:3neutoffdiagFLUX}
R_{\nu_e} &=& (P[\nu_e \rightarrow \nu_e] \Phi_{\nu_e} + P[\nu_{\mu} \rightarrow \nu_e] \Phi_{\nu_{\mu}} \nonumber \\ &+& P[\nu_{\tau} \rightarrow \nu_e] \Phi_{\nu_{\tau}})/\Phi_{\rm{tot}}, \nonumber \\
R_{\nu_{\mu}} &=& (P[\nu_e \rightarrow \nu_{\mu}] \Phi_{\nu_e} + P[\nu_{\mu} \rightarrow \nu_{\mu}] \Phi_{\nu_{\mu}} \nonumber \\ &+& P[\nu_{\tau} \rightarrow \nu_{\mu}] \Phi_{\nu_{\tau}})/\Phi_{\rm{tot}}, \nonumber \\
R_{\nu_{\tau}} &=& (P[\nu_e \rightarrow \nu_{\tau}] \Phi_{\nu_e} +
P[\nu_{\mu} \rightarrow \nu_{\tau}] \Phi_{\nu_{\mu}} \nonumber \\
&+& P[\nu_{\tau} \rightarrow \nu_{\tau}]
\Phi_{\nu_{\tau}})/\Phi_{\rm{tot}}
\end{eqnarray}
and so we find
\begin{eqnarray}
\label{eqn:3neutoffdiagFLUXP} \nonumber
R_{\nu_{e}}&=&0.264+0.300\varepsilon-a E^{n+1}[0.593\varepsilon-0.154]\times10^{9n+13},\\
\nonumber R_{\nu_{\mu}}&=&0.365-0.101\varepsilon+a E^{n+1}[0.167\varepsilon-0.013]\times10^{9n+13},\\
R_{\nu_{\tau}}&=&0.367-0.187\varepsilon+a E^{n+1}[0.409\varepsilon-0.116]\times10^{9n+13}.
\end{eqnarray}
Again, this result is only valid near the threshold of LV effects. We have also calculated the ratios in the large $a$ limit and find: $R_{\nu_e}:R_{\nu_{\mu}}:R_{\nu_{\tau}} \approx 0.42:0.57:0.013$ for $\varepsilon =1/3$ and $R_{\nu_e}:R_{\nu_{\mu}}:R_{\nu_{\tau}} \approx 0.70:0.27:0.027$ for $\varepsilon =1$. Our numerical calculations indicate that the transition between standard oscillation phenomenology and the phenomenology of the large $a$ limit takes place suddenly. If we assume what is perhaps the most natural choice of $a=M^{-1}_{\rm{Pl}}$ ($n=1$) or $a=M^{-2}_{\rm{Pl}}$ ($n=2$), the thresholds for these effects take place at $\sim 1$ TeV and $\sim 10$ PeV, respectively. Approximate numerical results are illustrated in figure~\ref{lvfig}.

\begin{figure}[tb]
    \includegraphics[scale=0.4,angle=90]{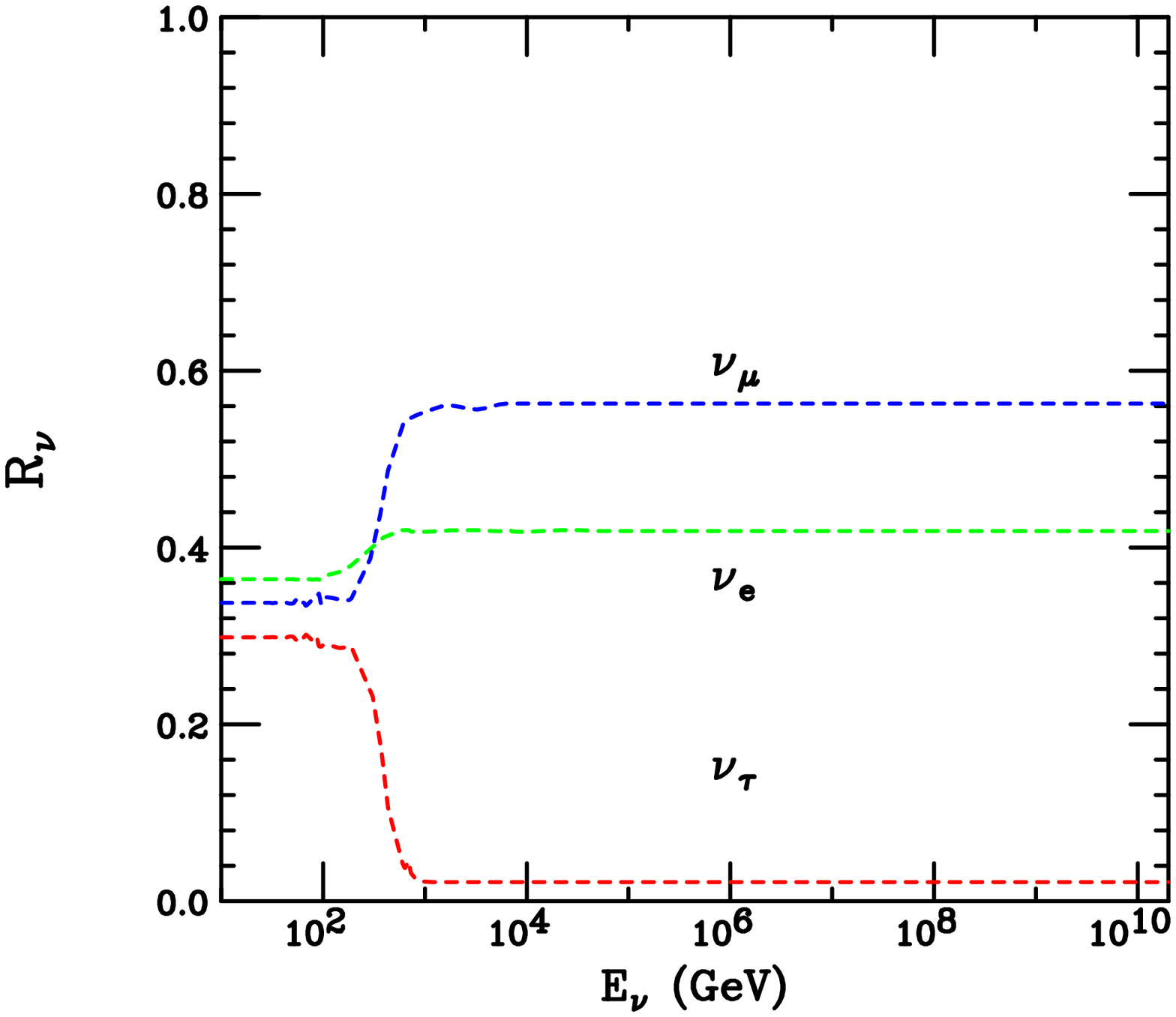} \hspace{0.7cm}
    \includegraphics[scale=0.4,angle=90]{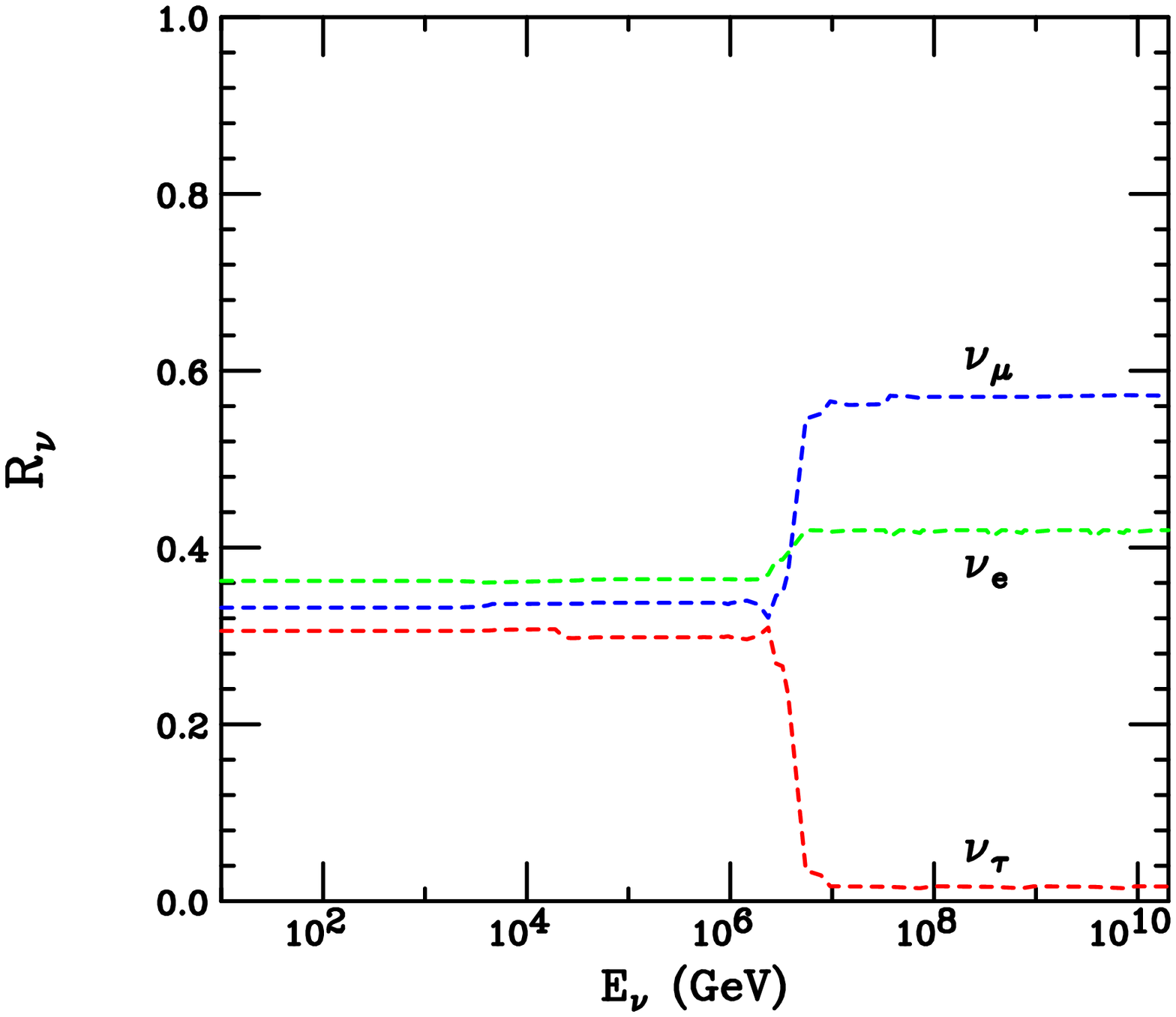} \\ \vspace{0.5cm}
    \includegraphics[scale=0.4,angle=90]{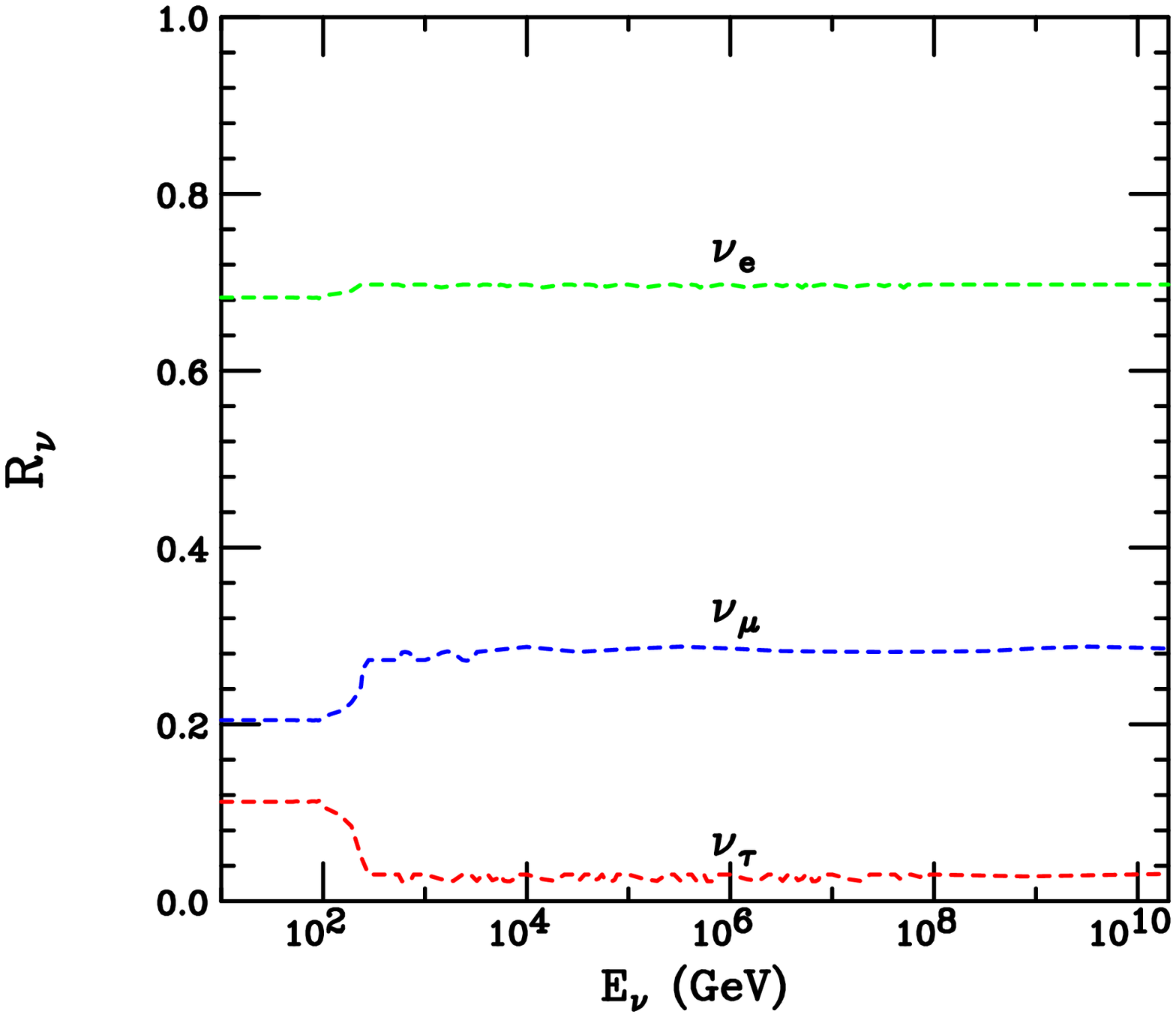} \hspace{0.7cm}
    \includegraphics[scale=0.4,angle=90]{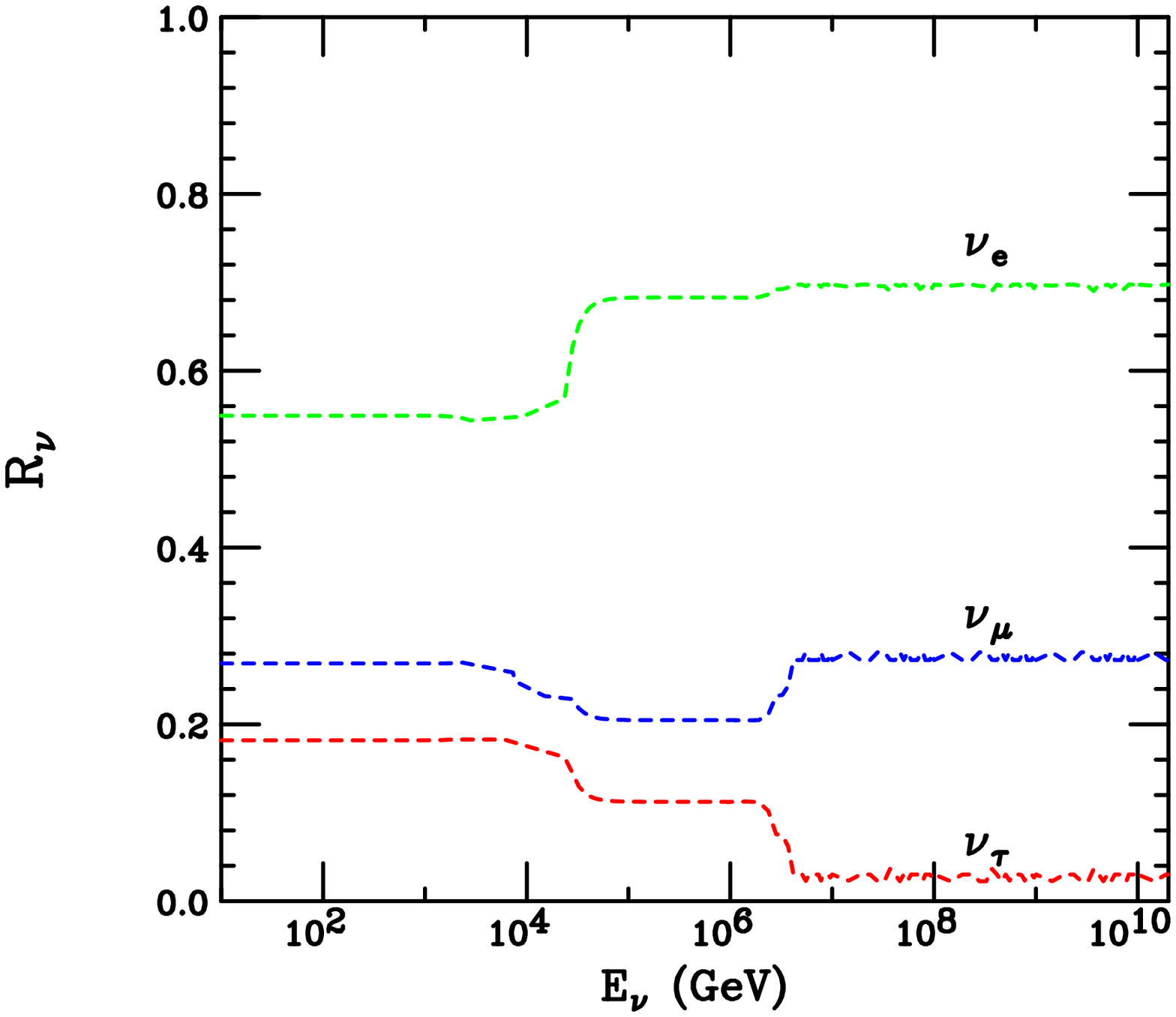} 
    \caption{The estimated effects of Lorentz violation on the ratio of neutrino flavors observed after propagation. The top frames correspond to neutrinos produced through pion decay, while the lower frames correspond to  (anti-)neutrinos produced through the decay of neutrons. The left and right frames display the behavior of models with Lorentz violating parameters proportional to $E^2$ and $E^3$, respectively.}
\label{lvfig} 
\end{figure}

\subsection{Quantum decoherence and neutrino oscillations}
We now turn to the violation of CPT without explicit Lorentz
violation and consider how quantum decoherence affects neutrino
oscillations. As we discussed in the previous section, quantum
decoherence causes the time evolution of the density matrix to be
altered:
\begin{equation}
\label{eqn:decoTEDM} \frac{d\rho}{dt}=-i[H,\rho]+\delta
H\!\!\,\!\!\!\slash\rho
\end{equation}
where $\delta H\!\!\,\!\!\!\slash$ arises due to the loss of
coherence. Considering two neutrinos  and expressing this equation
in the Pauli matrices basis, we find the time evolution of the
density matrix to take the form
\begin{equation}
\label{eqn:decoTEH}
\frac{d\rho_{\mu}}{dt}=(h_{\mu\nu}+h'_{\mu\nu})\rho_{\nu}
\end{equation}
where $h$ represents standard oscillations and $h'$ the
decoherence effects. Following Ref.~\cite{ellis}, we parameterize
$h'$ as
\begin{equation}
\label{eqn:decoPARAM} h'=-2\left(%
\begin{array}{cccc}
  0 & 0 & 0 & 0 \\
  0 & a & b & d \\
  0 & b & \alpha & \beta \\
  0 & d & \beta & \delta \\
\end{array}%
\right).
\end{equation}
The first row and column contain only zeros in order to conserve
probability and obey the second law of thermodynamics. We may
further constrain the theory if we assume that energy is conserved
within the neutrino system. In order for this to be the case, the
parameters $d,~\beta,~\delta$ must all be identically zero.
However, it is not clear what assumptions we should make from a
quantum gravity perspective and so we will also examine the
possibilities that these parameters have non-zero values. Using
Eqs.~(\ref{eqn:decoTEH}) and (\ref{eqn:decoPARAM}), we obtain
the equations,
\begin{eqnarray}
\dot{\rho}_{0} &=& 0,
\nonumber \\
\dot{\rho}_{1} &=& -2a \rho_{1} - 2\left( b- \frac{\Delta
m^{2}}{4E} \right) \rho_{2} -2 d \rho _{3},
\nonumber \\
\dot{\rho}_{2} &=& -2 \left( b+ \frac{\Delta m^{2}}{4E}
\right)\rho_{1} -2\alpha \rho_{2} -2 \beta \rho _{3},
\nonumber \\
\dot{\rho}_{3} &=& -2d \rho _{1} -2\beta \rho _{2} -2 \delta \rho
_{3}. \label{diffeqn:rho}
\end{eqnarray}
In order to find the oscillation probability, we solve these
equations with suitable initial conditions. There does not exist,
however, a simple closed form for the solution to these equations
and so, for illustrative purposes, we consider two limiting cases.
The first is that the parameters $d$ and $\beta$ are zero and the
second is that all the decoherence parameters are zero
\textit{except} $d$. In the first case, the oscillation
probability takes the form
\begin{eqnarray}
\label{eqn:decoP1} \nonumber
P[\nu_{\alpha}\rightarrow\nu_{\beta}]&=&
\frac{1}{2}\left\{\cos^{2}2\theta(1-e^{-2\delta
L})\right.\\
&&\!\!\!\!\!\!\!\!\!\!\!\!\!\!\!\!\!\!\!\!\!+\left.\sin^{2}2\theta\left[1-e^{-(a+\alpha)L}\cos\left(2L\left[\left(\frac{\Delta
m^{2}}{4E}\right)^{2}-\frac{1}{4}(\alpha-a)^{2}-b^{2}\right]^{\frac{1}{2}}\right)\right]\right\}.
\end{eqnarray}
If we assume complete positivity \cite{benatti}, necessary for the
Lindblad form, (as described in section \ref{sec:VioLVCPT}) with
energy conservation, we find the simplest possible extension to
standard neutrino oscillations which includes dehoherence. In this
case $a=\alpha$ with all other parameters equal to zero.
Considering the second approximation, with non-zero $d$ only, gives
the probability of oscillation to be
\begin{eqnarray}
\label{eqn:decoP2} \nonumber
P[\nu_{\alpha}\rightarrow\nu_{\beta}]&=&\frac{1}{2}\left\{\cos^{2}2\theta\left[1-\frac{\omega^{2}}{\Omega_{d}^{2}}+\frac{d^{2}}{\Omega_{d}^{2}}\cos(2\Omega_{d}L)\right]\right.\\
\nonumber &&+\left.\sin^{2}2\theta\left[1+\frac{d^{2}-\omega^{2}}{\Omega_{d}^{2}}\cos(2\Omega_{d} L)\right]\right.\\
&&+\left.\sin4\theta\left[\frac{d}{\Omega_{d}}\sin(2\Omega_{d}L)\right]\right\},
\end{eqnarray}
where $\Omega_{d}=\sqrt{w^{2}+d^{2}}$ and $\omega=\Delta
m^{2}/4E$. Note the similarity to the probability given in
Eq.~(\ref{eqn:2neutnondiagOP}).
\par
So far, we have said nothing about the form of the decoherence
parameters. It is clear that the decoherence parameters must have
dimensions of energy but it is also possible that they have an explicit
dependence on the neutrino energy. In the literature
\cite{decolit1,decolit2,decolit3}, three models have received
significant attention, specifically, with energy dependences
$E^{0},~E^{-1}$ and $E^{2}$. From a quantum gravity
perspective, an energy dependence of $E^{2}$ is particularly
interesting \cite{decostring1,decostring2}.
\par
In a similar way to the LV case, it is also worthwhile studying
the three neutrino case. Again, the situation is made somewhat
more complicated by the existence of 3 mixing angles, two
independent mass differences and many decoherence parameters. We
follow an analogous method to that described above but now,
instead of using the Pauli matrices as a basis, we choose the
generators of SU(3). In order to derive analytical results, we
choose the form of the additional decoherence matrix so we may
solve the differential equations straightforwardly but keep as
many decoherence parameters as possible. Including standard
oscillations, we therefore find the analogous sum of $h+h'$ in
Eq.~(\ref{eqn:decoTEH}) to be given by
\begin{equation}
\label{eqn:3neutdecoH} \mathcal{H} = \left(%
\begin{array}{ccccccccc}
  0 & 0 & 0 & 0 & 0 & 0 & 0 & 0 & 0 \\
  0 & A & B+\omega_{21} & 0 & 0 & 0 & 0 & 0 & 0 \\
  0 & B-\omega_{21} & \Lambda &  0& 0 & 0 & 0 & 0 & 0 \\
  0 & 0 & 0 & \Psi & 0 & 0 & 0 & 0 & 0 \\
  0 & 0 & 0 & 0 & x & y+\omega_{31} & 0 & 0 & 0 \\
  0 & 0 & 0 & 0 & y-\omega_{31} & z & 0 & 0 & 0 \\
  0 & 0 & 0 & 0 & 0 & 0 & a & b+\omega_{32} & 0 \\
  0 & 0 & 0 & 0 & 0 & 0 & b-\omega_{32} & \alpha & 0 \\
  0 & 0 & 0 & 0 & 0 & 0 & 0 & 0 & \delta \\
\end{array}%
\right),
\end{equation}
where $\omega_{ji}=\Delta m^{2}_{ji}/4E$. Solving these equations
and using Eq.~(\ref{eqn:2neutnondiagP}) gives the oscillation
probability from flavor $p$ to $q$ as
\begin{eqnarray}
\label{eqn:3neutdecoP} \nonumber P[\nu_{p}\rightarrow \nu_{q}] &=&
\frac{1}{3}+\frac{1}{2}(U_{p1}^{2}-U_{p2}^{2})(U_{q1}^{2}-U_{q2}^{2})e^{-2\Psi
L}\\
\nonumber
&&+\frac{1}{6}(U_{p1}^{2}+U_{p2}^{2}-2U_{p3}^{2})(U_{q1}^{2}+U_{q2}^{2}-2U_{q3}^{2})e^{-2\delta
L}\\
\nonumber &&+2U_{p1}U_{p2}U_{q1}U_{q2}e^{-(A+\Lambda)L}\left[\cos(2\Omega_{21}L)+\frac{(\Lambda-A)}{2\Omega_{21}}\sin(2\Omega_{21}L)\right]\\
\nonumber &&+2U_{p1}U_{p3}U_{q1}U_{q3}e^{-(x+z)L}\left[\cos(2\Omega_{31}L)+\frac{(z-x)}{2\Omega_{31}}\sin(2\Omega_{31}L)\right]\\
&&+2U_{p2}U_{p3}U_{q2}U_{q3}e^{-(a+\alpha)L}\left[\cos(2\Omega_{32}L)+\frac{(\alpha-a)}{2\Omega_{32}}\sin(2\Omega_{32}L)\right],
\end{eqnarray}
where the $U$'s denote the entries in the standard mixing matrix
introduced in Eq.~(\ref{eqn:3x3mix}). If we again consider
neutrinos travelling large distances from astrophysical objects,
we can average the $\sin$ and $\cos$ terms to zero. If, for simplicity, we assume $\Psi=\delta$, the probability reduces to
\begin{eqnarray}
\label{eqn:3neutdecoPsimp} \nonumber P[\nu_{p}\rightarrow \nu_{q}]
&=& \frac{1}{3}+\frac{1}{6}e^{-2\delta
L}\left[3(U_{p1}^{2}-U_{p2}^{2})(U_{q1}^{2}-U_{q2}^{2})\right.\\
\nonumber
&&\left.+(U_{p1}^{2}+U_{p2}^{2}-2U_{p3}^{2})(U_{q1}^{2}+U_{q2}^{2}-2U_{q3}^{2})\right].
\end{eqnarray}
We can now derive equations describing the flavor composition at the
detector in the same way as we did in the LV case: 
\begin{eqnarray}
\label{eqn:3neutdecoR} \nonumber
R_{\nu_{e}}&=&\frac{1}{3}+e^{-2\delta
L}[0.287\varepsilon-0.065],\\
\nonumber R_{\nu_{\mu}}&=&\frac{1}{3}-e^{-2\delta
L}[0.096\varepsilon-0.03],\\
R_{\nu_{\tau}}&=&\frac{1}{3}-e^{-2\delta
L}[0.189\varepsilon-0.034].
\end{eqnarray}
\subsection{CPTV and the LSND anomaly}
Having discussed how LV and CPTV may alter the
phenomenology of neutrino oscillations, we are now in a position
to discuss these effects within the context of the anomalous LSND result. As we discussed in section~\ref{sec:Nuosc}, the LSND
experiment found a mass difference for anti-neutrinos which is
incompatible with that of standard oscillations with three
neutrinos. Assuming that these results are indeed correct, there are two ways to explain this result. The first is
to assume that there are more than three neutrinos, with the
additional neutrino(s) being sterile and not directly detectable. The second is to violate CPT invariance.
\subsubsection{Direct CPTV and the LSND anomaly}
In order to reconcile the LSND result with the rest of the
neutrino oscillation data, it seems necessary to modify the
neutrino sector to include different independent mass splittings
for neutrinos and anti-neutrinos. In this way, we preserve the
number of neutrinos but allow the mass differences to differ
between the neutrino and anti-neutrino sectors. The present
situation, taking in to account solar, atmospheric and
KamLAND data, disfavours this scenario \cite{global}, however, in both two
 and three neutrino models \cite{lsnd2,lsnd3}, leaving room only for CPTV in four neutrino models where the mixing parameters
may be different in neutrino and anti-neutrino sectors \cite{lsnd4}.
\subsubsection{CPTV from quantum decoherence and the LSND anomaly}
A second possibility for explaining the LSND result without
enlarging the neutrino sector is to consider quantum decoherence
in the anti-neutrino sector only \cite{lsndnick}. The oscillation
probability for three anti-neutrinos now takes the form of
Eq.~(\ref{eqn:3neutdecoP}), whilst neutrinos experience no
quantum decoherence. Since decoherence parameters are present in
the arguments of the sine and cosine terms, they alter the
effective mass differences leading to an apparent difference
between the measured mass differences in the neutrino and
anti-neutrino sectors. It is therefore possible to reconcile the
LSND results with other existing oscillation data \cite{lsndnick}.
However, this particular model fails to fit the spectral
distortions observed in the KamLAND experiment \cite{KLspec}.
Having said that, the authors of Ref.~\cite{lsndnick} chose only one set of quantum decoherence parameters and so there is still much scope for
further investigation.

\section{Sources and Detection of High-Energy Neutrinos}
\label{mysec}

\subsection{Sources of high-energy neutrinos}

High-energy neutrinos are thought to be generated in a wide range of astrophysical sources. Such sources may include Active Galactic Nuclei (AGN) \cite{agn}, Gamma-Ray Bursts (GRB) \cite{grb}, microquasars \cite{microquasars}, supernova remnants, star clusters and X-ray binaries. Also, ultra-high protons or nuclei travelling over cosmological distances can interact with the Cosmic Microwave Background (CMB) and/or Cosmic Infra-Red Background (CIRB), generating what is often called the cosmogenic neutrino flux \cite{cosmogenic}.

In any of these sources, there are basically two mechanisms by which high-energy neutrinos are generated. Firstly, Fermi accelerated protons (or charged nuclei) can collide with hadronic or photonic targets generating charged and neutral pions. These charged pions then decay, $\pi^{+} \rightarrow \mu^+ \nu_{\mu} \rightarrow e^+ \bar{\nu}_e \nu_e \nu_{\mu}$, $\pi^{-} \rightarrow \mu^- \bar{\nu}_{\mu} \rightarrow e^- \nu_e \bar{\nu}_e \bar{\nu}_{\mu}$, generating electron and muon neutrinos and anti-neutrinos. Secondly, atomic nuclei which undergo Fermi acceleration can be disintegrated by interacting with infra-red photons surrounding their source. Neutrons broken off of such a nucleus can then decay, $n \rightarrow p^+ e^- \bar{\nu}_e$, generating a flux of electron anti-neutrinos.


The important thing to keep in mind regarding these two mechanisms for high-energy neutrino generation is the quantity of neutrinos produced of various flavors. Neutrinos produced in charged pion decay follow the ratio: $\nu_e:\nu_{\mu}:\nu_{\tau} = 1/3:2/3:0$, while those produced in neutron decay follow: $\nu_e:\nu_{\mu}:\nu_{\tau} = 1:0:0$. It is also important to note that pion decay can generate both neutrinos and anti-neutrinos, while neutron decay generates only anti-neutrinos.

For the purposes of the measurement of flavor ratios, identified point sources of high-energy neutrinos are considerably more useful than diffuse fluxes. With such a source (or sum of sources), the distance the neutrinos have propagated will likely be known. Furthermore, the background from atmospheric neutrinos can be controlled by only considering events from one direction in the sky. In some sources which emit neutrinos only for short lengths of time, GRB and AGN flares for example, the background can be further reduced by only considering events in particular time windows. For these reasons, GRB and AGN are likely to be among the most useful source of high-energy neutrinos for the purposes of flavor identification although bright and nearby (galactic) sources, if present, could also be very useful. 

In addition to these theoretical arguments, there is some limited experimental evidence that might suggest the existance of bright point sources of high-energy neutrinos. Firstly, it has been argued recently that anisotropies observed in the cosmic ray spectrum at EeV energies is the result of neutrons propagating from galactic sources. If this is the case, then large fluxes of high-energy (anti-)neutrinos will also be generated \cite{neutrondecay}. Secondly, the AMANDA-II experiment has recently reported the detection of two neutrinos coincident with TeV flares seen by the Whipple gamma-ray telescope from the blazar (AGN) 1ES 1959+650 \cite{1959}. These events were not found in a blind analysis, however, so their statistical significance cannot be determined. If these neutrinos are the product of this TeV blazar, it would suggest the existance of very bright point sources of high-energy neutrinos.

More exotic processes which do not fall into these two descriptions may also be capable of generating high-energy cosmic neutrinos. Such possibilities include annihilating or decaying dark matter or topological defects \cite{darkmatter}, Hawking radiating primordial black holes \cite{hawkbh}, or the interactions of ultra-high energy neutrinos with the cosmic neutrino background via the $Z$-burst mechanism \cite{zburst}. For a review of sources of high-energy neutrinos and other aspects of high-energy neutrino astronomy, see Ref.~\cite{neutrinoreview}.

\subsection{High-energy neutrino detection}

Once such neutrinos are generated and propagate to Earth, they can be detected in one of several ways. Neutral current interactions of neutrinos of all flavors with nucleons generates hadronic showers which can be observed. Charged current interactions of electron and muon neutrinos generate, in addition to hadronic showers, potentially observable electromagnetic showers and muons, respectively. The tau leptons generated in the charged current interactions of tau neutrinos can produce a class of events unique to tau neutrinos: double bangs and lollipops.

Many of the experimental techniques being developed and deployed are only capable of detecting showers generated in high-energy neutrino interactions. Observing shower events alone will not enable the flavor ratios of a flux of cosmic neutrinos to be identified, however. For this reason we focus on experiments capable of observing high-energy neutrinos in the form of showers, muon tracks and tau-unique events. In particular, we focus on next generation, kilometer-scale, optical Cerenkov detectors. These include IceCube, currently under construction at the South Pole, and possibly a future kilometer-scale neutrino telescope in the Mediterranean Sea, sometimes called KM3.

\subsubsection{Shower events}

All high-energy neutrino interactions produce an electromagnetic and/or hadronic shower. The probability of detecting a hadronic shower produced in a neutral current interaction as a neutrino travels through the effective area of the detector is given by
\begin{equation}
P_{\nu \rightarrow \rm{shower}} = \rho N_A L \int^1_{E^{\rm{thr}}_{\rm{sh}}/E_{\nu}} \frac{d \sigma}{dy} dy,
\end{equation}
where $\rho$ is the target nucleon density, $N_A$ is Avogadro's number, $L \approx 1$ km is the length of the detector, $d\sigma/dy$ is the differential neutrino-nucleon neutral current cross section \cite{crosssection}, $y$ is the fraction of energy which is transferred from the neutrino (and therefore the fraction of energy which goes into the shower) and $E^{\rm{thr}}_{\rm{sh}} \approx 3$ TeV is the threshold energy for the experiment detecting a shower event.

In the charged current interactions of electron neutrinos, all of the neutrino's energy goes into a combination of electromagnetic and hadronic showers. In this case, the probability of detecting a shower reduces to 
\begin{equation}
P_{\nu \rightarrow \rm{shower}} = \rho N_A L \sigma,
\end{equation}
if $E_{\nu} > E^{\rm{thr}}_{\rm{sh}}$, and zero otherwise. $\sigma$ in this expression is the total charged current neutrino-nucleon cross section \cite{crosssection}.

In these expressions, we have made the simplifying assumption that only showers generated inside of the detector volume can be detected. This is not always true, particularly at very high energies. Very energetic showers which are initiated outside of the experiment's instrumented volume can expand into the experiment. Another way of saying this is that the effective volume of such an experiment is generally larger than its instrumented volume at very high energies.

\subsubsection{Muon events}

Charged current interactions of high-energy muon neutrinos produce muons which can travel through the medium of the experiment (ice or water) and potentially into the detector volume. The energy loss rate of such a muon is given by
\begin{equation}
\frac{dE}{dX} \approx -\alpha - \beta E,
\end{equation}
where the parameters in ice or water are given by $\alpha \approx 2.0$ MeV cm$^2$/g and $\beta \approx 4.2 \times 10^{-6}$ cm$^2$/g \cite{dutta}. The distance a muon travels before its energy drops below the threshold, $E^{\rm{thr}}_{\mu}$, is given then by
\begin{equation}
R_{\mu} = \frac{1}{\beta} \ln\bigg[\frac{\alpha + \beta E_{\mu}}{\alpha + \beta E^{\rm{thr}}_{\mu}}\bigg].
\end{equation}
This quantity if often referred to as the muon range. $E^{\rm{thr}}_{\mu} \sim$ 50-100 GeV are typical for high-energy neutrino telescopes. The muon range can extend for many kilometers for very high energy muons, dramatically increasing the number of muon tracks that are observed.

The probability of detecting a muon produced in a charged current interaction is given by
\begin{equation}
P_{\nu_{\mu} \rightarrow \mu} \approx \rho N_A \sigma R_{\mu},
\end{equation}
where, here, $\sigma$ is the total charged current neutrino-nucleon cross section \cite{crosssection}. In cases with the muon neutrino comes from above the ice or water, the value of $R_{\mu}$ used in this equation should not exceed the depth of the experiment. Similarly, if the neutrino comes from below the detector, the rock or other material below the ice or water should be accounted for.

\subsubsection{Events involving tau neutrinos}

For tau neutrinos with energies less than $\sim$PeV, their interactions generate only shower events, as any tau leptons generated decay before they can be identified. Higher energy tau neutrinos, on the other hand, can be identified by the combined signatures of tau tracks and showers.

Double bang \cite{doublebang,measure} events are produced with a tau neutrino interacts via charged current inside of the detector, producing a tau lepton which travels across the detector to decay, still inside of the detector volume. The hadronic shower produced in the initial interaction constitutes the first ``bang'' while the second shower is generated in the tau's decay. A tau lepton can travel a distance
\begin{equation}
R_{\tau} = \frac{E_{\tau}\, c \tau_{\tau}}{m_{\tau}} \sim \frac{E_{\tau}}{1 \, \rm{PeV}}  \, 50 \, \rm{meters}
\end{equation}
before decaying. To resolve two showers in a high-energy neutrino telescope, they must be separated by a distance of roughly 100-400 meters. Furthermore, they cannot be separated by more than about 1000 meters and both be within the detector volume. Therefore double bang events are most useful in the energy range between a few PeV and one hundred PeV.

At higher energies, lollipop events become very useful \cite{measure}. A lollipop event is observed when the first shower of a double bang event occurs outside of the detector (without being observed) with the tau track extending into the detector and decaying. You might also think that an observation of the first shower with a tau track would be a useful signature, but muons and taus produced in ordinary hadronic or electromagnetic showers could mimic such an event.

The probability of observing a double bang event from a given incident tau neutrino is given by
\begin{equation}
P_{\rm{DB}} \approx \rho N_A \int^1_0 dy \frac{d \sigma}{dy} \int^L_{x_{\rm{min}}} dx \frac{(L-x)}{R_{\tau}} e^{-x/R_{\tau}}, 
\end{equation}
where as the probability for a lollipop event is
\begin{equation}
P_{\rm{LP}} \approx \rho N_A (L-x_{\rm{min}}) \int^1_0 dy \frac{d \sigma}{dy} e^{-x/R_{\tau}}. 
\end{equation}
In each of these expressions, $x_{\rm{min}} \sim 100-400$ meters, is the minimum shower seperation required to seperate the showers.

For a detailed description of both double bang and lollipop tau events, see Ref.~\cite{measure}.

\subsubsection{Neutrino events in cosmic ray experiments}

Although in this article we focus on next generation optical Cerenkov neutrino telescopes, it is interesting to point out that ultra-high energy cosmic ray experiments may also have a limited ability to resolve neutrino flavors. These experiments can detect ultra-high energy neutrinos in essentially two ways. First, quasi-horizontal neutrinos can penetrate deeply into the atmosphere before interacting, producing showers which are distinguishable from those initiated by cosmic rays. Second, tau neutrinos which skim the Earth can, through charged current interactions, produce tau leptons which escape the Earth before decaying and generating a shower \cite{neutrinocr}. While the former signature can be produced by neutrinos of all three flavors, the latter are uniquely generated by tau neutrinos. This, in principle, could be used to measure the fraction of the ultra-high energy neutrino flux which consists of tau neutrinos.

\section{Signatures of CPT and Lorentz Violation}

There are two primary methods of probing physics beyond the Standard Model with high-energy neutrinos. First, neutrino-nucleon interactions can be observed measuring the cross section \cite{nucross} and other characteristics in hope of identifying new interactions: microscopic black hole or P-brane production \cite{nublackhole}, processes resulting from low-scale gravity \cite{nugravity}, string effects \cite{nustring} or electoweak instantons \cite{nuinstanton}. Second, the ratios of neutrino flavors can be measured, potentially identifying the effects of neutrino decay \cite{nudecay}, pseudo-Dirac states \cite{nupseudodirac} or quantum decoherence \cite{decolit1}. In this article, we focus on using this latter technique to constrain or discover the effects of CPT and Lorentz violation.

\subsection{Flavor ratio predictions}

To assess the prospects of identifying the violation of CPT and Lorentz invariance in the high-energy neutrino sector, we must first determine the flavor ratios predicted in various scenarios. First of all, we consider the case with only known physics, and no CPT or Lorentz violation. For an initial set of ratios corresponding to neutrinos from pion decay, $\nu_e:\nu_{\mu}:\nu_{\tau} = 1/3:2/3:0$, after oscillations over a long distance (which will always be the case for high-energy neutrino astronomy), these ratios become  $\nu_e:\nu_{\mu}:\nu_{\tau} \approx 0.36:0.33:0.30$. (Anti)-neutrinos coming from neutron decay, on the other hand, are generated in the ratio $\nu_e:\nu_{\mu}:\nu_{\tau} = 1:0:0$, which oscillates to $\nu_e:\nu_{\mu}:\nu_{\tau} \approx 0.56:0.26:0.18$.

These ratios can be changed dramatically if the effects of CPT or Lorentz violation are present. In the case of LV, above the energy threshold for such effects, the neutrino flavor ratios from pion decay are modified as $\nu_e:\nu_{\mu}:\nu_{\tau} = 1/3:2/3:0 \rightarrow 0.42:0.57:0.013$. The flavor ratios from neutron decay are modified as $\nu_e:\nu_{\mu}:\nu_{\tau} = 1:0:0 \rightarrow 0.70:0.27:0.027$. In the case of quantum decoherence, on the other hand, all ratios shift toward $\nu_e:\nu_{\mu}:\nu_{\tau} = 1/3:1/3:1/3$, regardless of their source (see figure~\ref{qdc}).

\subsection{Lorentz Violation}

There are two potentially identifiable features in the neutrino flavor ratios predicted as a result of LV. First, in the case of neutrinos generated via pion decay, a particularly large fraction of these neutrinos will be of muon flavor. Secondly, regardless of whether the neutrinos are generated in pion or neutron decay, only a very small fraction will appear with tau flavor.

\subsubsection{A large muon neutrino fraction}

If the effects of LV are considered in neutrinos generated via pion decay, the result can be a neutrino flux which is of nearly 60\% muon flavor. A large muon to shower ratio could, therefore, be seen as a signature of LV. The ability of high-energy neutrino telescopes to measure the ratio of muon to shower events and corresponding flavor ratio is discussed in detail in Ref.~\cite{measure}.

\subsubsection{A tau neutrino deficit}

Looking for a deficit of high-energy cosmic tau neutrinos may also be useful in identifying the effects of LV. Events that are identifiable as tau neutrinos (double bang and lollipop events) are somewhat rare, however. Assuming a spectrum proportional to $E_{\nu}^{-2}$, a flux of $E^2_{\nu_{\tau}} dN_{\nu_{\tau}}/dE_{\nu_{\tau}} = 10^{-7}$ GeV cm$^{-2}$ s$^{-1}$ would yield only about 0.35 tau-unique events per year in a cubic kilometer experiment. Over several years of observation, observing a deficit of tau neutrinos may indeed become possible if bright point sources of PeV-EeV neutrinos are discovered.

The absence of tau neutrino induced Earth-skimming showers at cosmic ray experiments, such as Auger, could also be an anticipated signature of LV effects. As experiments such as Auger most efficiently detect showers with energies above $\sim 10^{8}$ GeV, or so, they are particularly well suited for testing the effects of LV.

\subsection{Quantum Decoherence}

\begin{figure}[tb]
    \includegraphics[scale=0.45,angle=90]{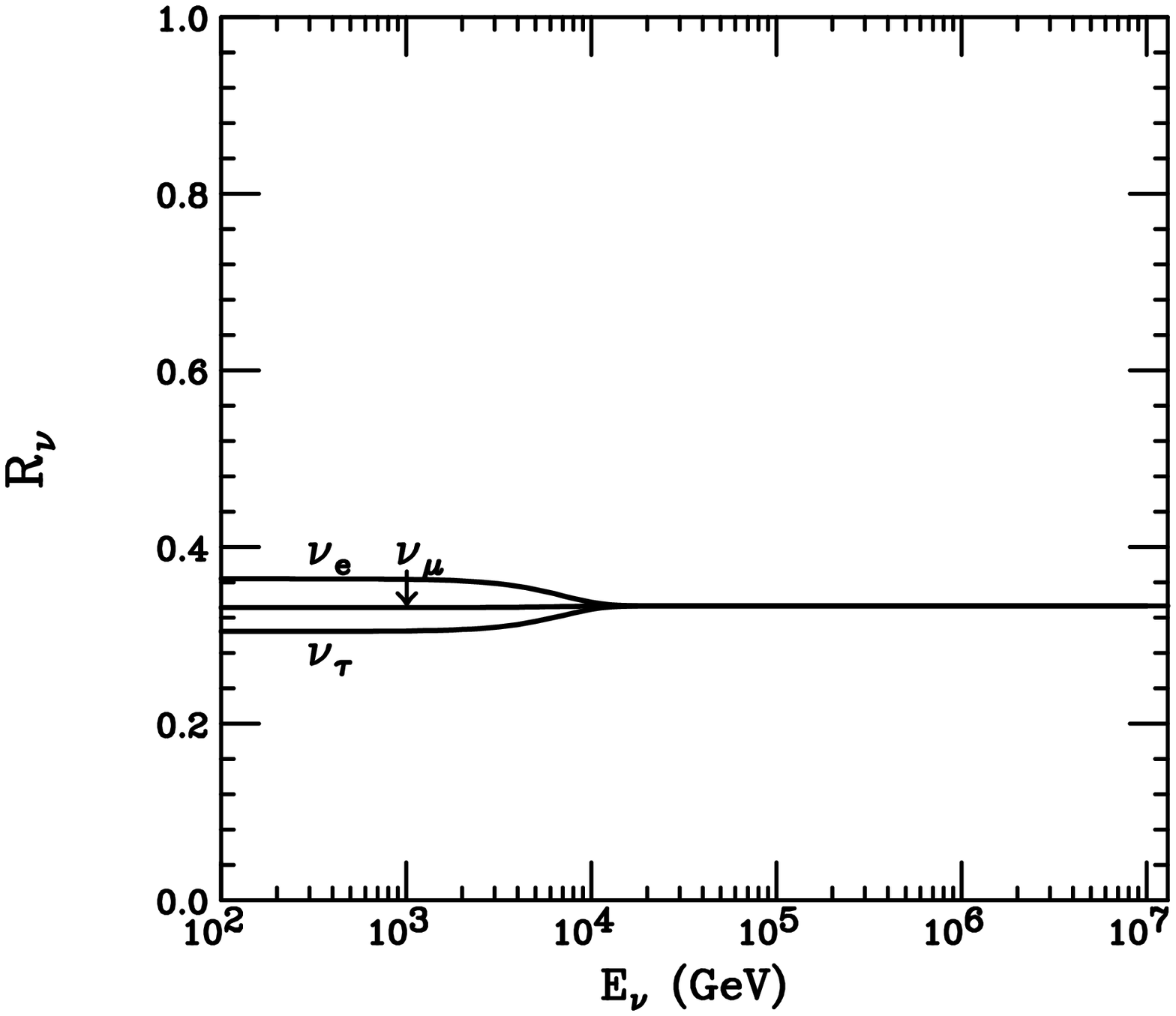}
    \includegraphics[scale=0.45,angle=90]{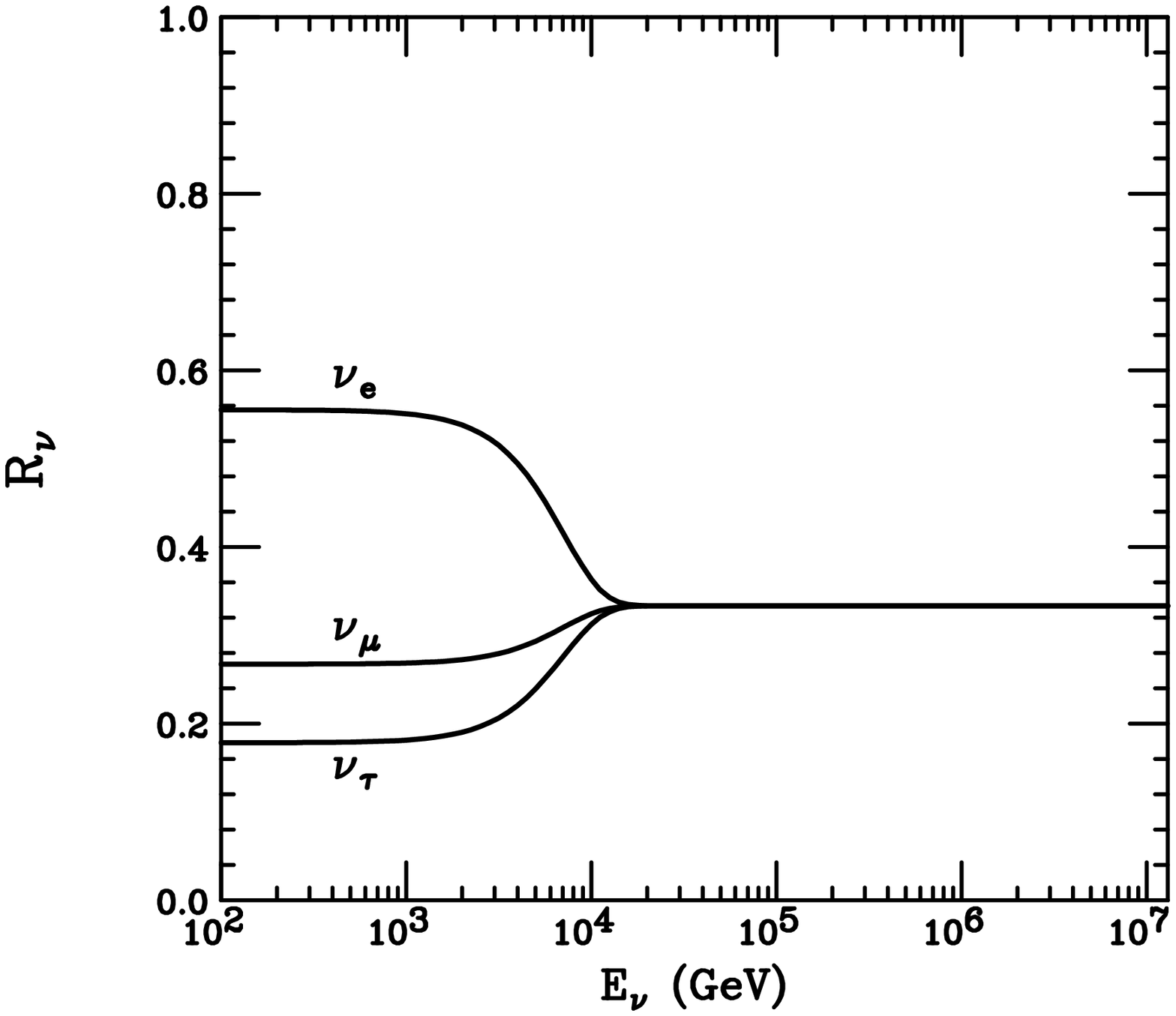}
    \caption{The effect of quantum decoherence on the ratios of neutrino flavors. In the left frame, the initial ratios are set to those values found from pion decay ($\nu_e:\nu_{\mu}:\nu_{\tau}=1/3:2/3:0$) while in the rigth frame the ratios from neutron decay  ($\nu_e:\nu_{\mu}:\nu_{\tau}=1:0:0$) are shown. In both cases, we have considered a model with $\delta L = (E/10,000 \, \rm{GeV})^2$. In the case of neutrinos generated in pion decay, quantum decoherence has little impact on the flavor ratios (left frame). The flavor ratios of (anti-)neutrinos generated in neutron decay, on the other hand, can be dramatically effected (right frame).}
\label{qdc}
\end{figure}

\begin{figure}[tb]
    \includegraphics[scale=0.45,angle=90]{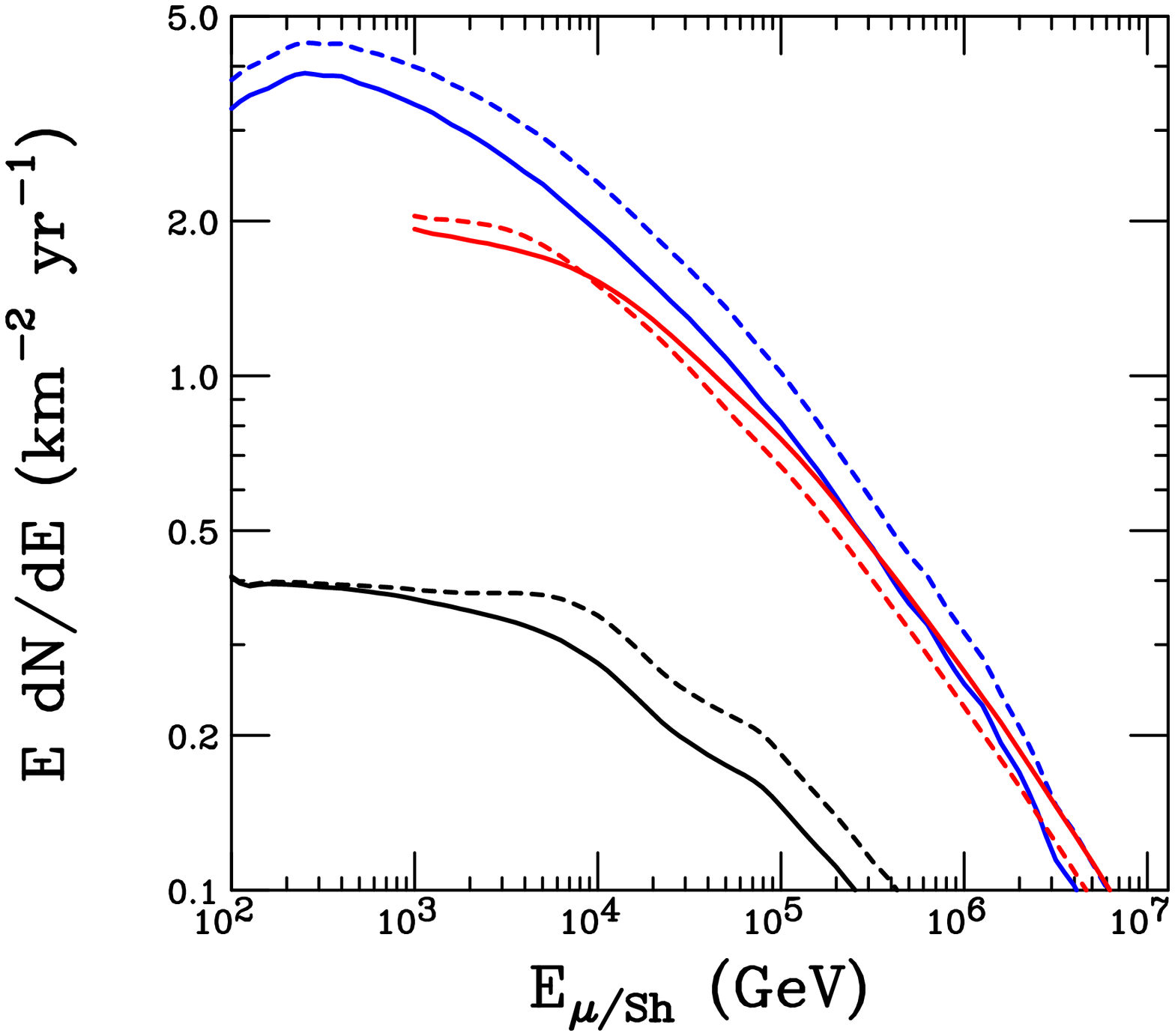}
    \includegraphics[scale=0.45,angle=90]{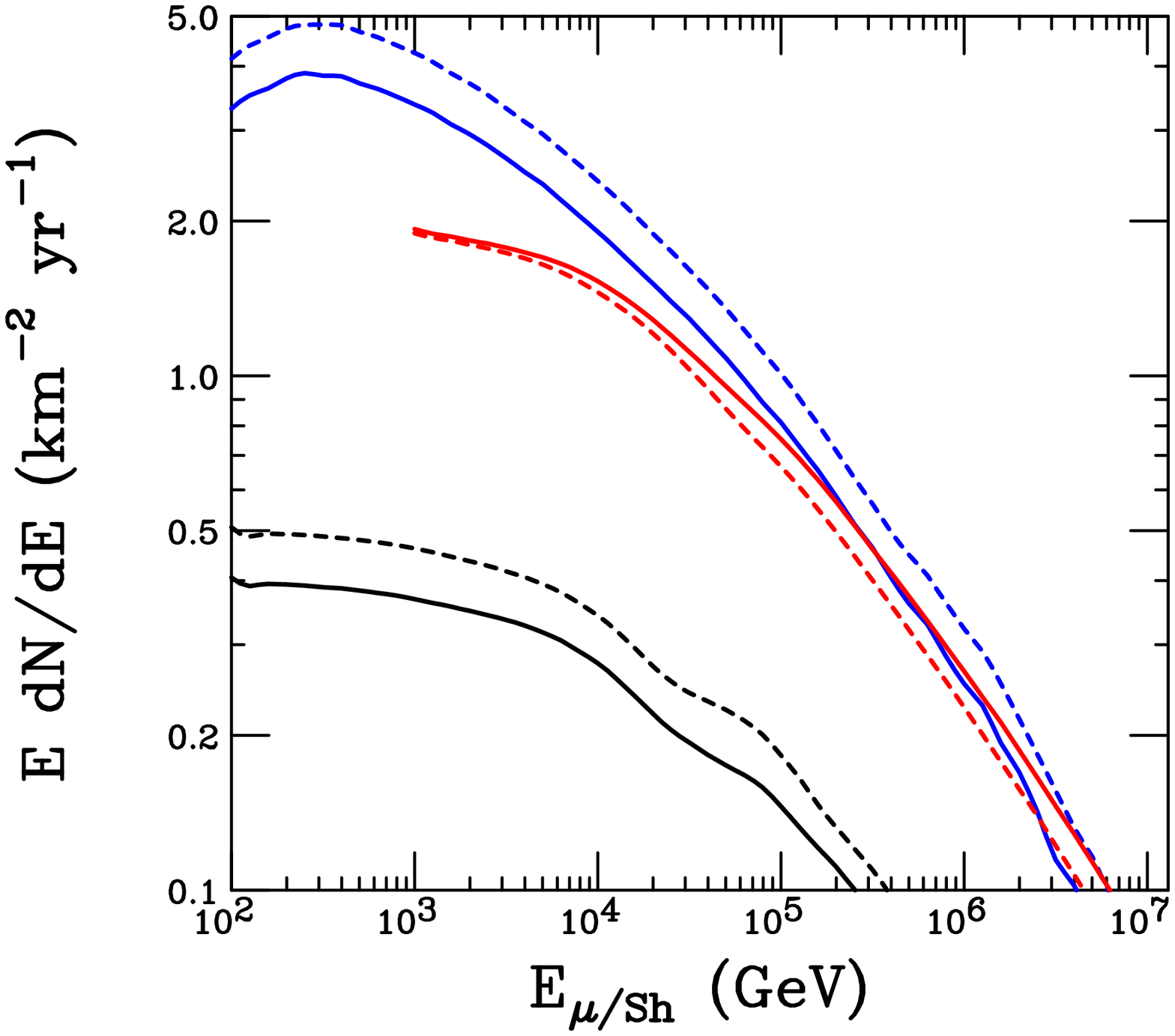}
    \caption{The effect of quantum decoherence on the events observed in a next generation high-energy neutrino telescope, such as IceCube or KM3. Solid lines are the predicted distribution of events with no effects of quantum decoherence, while the dashed lines represent the cases of $\delta L = (E/10,000 \, \rm{GeV})^2$ (left frame) and full decoherence (right frame). The (from top to bottom on the left side of the frames) blue, red and black lines show the distributions of muon events, electromagnetic/hadronic shower events, and all muon events with a vertex contained in the detector volume, respectively. The flavor ratios used are for a source producing (anti-)neutrinos via neutron decay. The rates have been normalized to a total neutrino flux of $E^2_{\nu} dN_{\nu}/dE_{\nu} = 10^{-7}$ GeV cm$^{-2}$ s$^{-1}$.}
\label{qd}
\end{figure}

The signature of quantum decoherence in cosmic neutrinos, is the presence of an equal fraction of neutrinos of each flavor. Since sources which produce neutrinos through pion decay generate nearly this ratio (after oscillations) without the effect of quantum decoherence, they are not very useful in probing for these effects. Sources of high-energy (anti-)neutrinos produced through neutron decay, on the other hand, can be used to potentially identify these effects \cite{decolit1}.

In figure~\ref{qd}, we show the effects of quantum decoherence on the events observed in a next generation high-energy neutrino telescope. In the left frame, we consider a model with a $\delta \propto E^2$ dependence, normalized such that quantum decoherence sets in at the 10 TeV scale ($\delta = (E/10,000 \, \rm{GeV})^2/L$). In the right frame, the source is sufficiently distant that the effects of quantum decohence have set in fully ($\nu_e:\nu_{\mu}:\nu_{\tau} = 1/3:1/3:1/3$). In either case, the number of muon tracks detected are increased and the number of showers detected decreases. Since the normalization of this flux will likely be unknown, it is the ratio of these event types that is of the most interested to us.

In figure~\ref{qd}, we plot the total rate of muon events and the rate of contained muon events ({\it i.e.} with a vertex contained within the detector volume) seperately. We do this to illustrate that energy dependent effects, such as those shown in the left frame of this figure, are more clearly identified by using only contained events. The statistics are considerably better when using all of the events, however. Depending on the flux of high-energy neutrinos present, a strategy to use these different types of events most effectively will have to be devised.

The prospects for detecting the effects of quantum decoherence or other signatures of CPT or Lorentz violation ultimately depend on the variety of high-energy neutrino sources which exist in nature. A detailed study of IceCube's sensitivity to decoherence effects from observing anti-neutrinos from the Cygnus spiral arm is currently underway \cite{cygnus}. This is likely to be one of the most useful sources for constraining the effects of decoherence.

\section{Conclusions}
\label{conc}

High-energy neutrino astronomy provides an opportunity to observe particles at extremely high energies and over extremely long baselines. Both of these characteristics make such experiments particularly adept at testing for the effects of CPT and Lorentz violation.

In this article, we have discussed in detail the effects that CPT and Lorentz violation can have on the flavors of high-energy cosmic neutrinos observed at Earth. After discussing the theoretical basis and motivations for such effects, we calculated the ratios of neutrino flavors predicted to be observed from astrophysical sources of high-energy neutrinos in various CPT and Lorentz violating scenarios. 

The effects of Lorentz violation may potentially be detected or constrained by the observation, or lack thereof, of anomalously large fractions of the cosmic neutrino spectrum consisting of muon neutrinos, accompanied by very few tau neutrinos. This deviation from the standard neutrino oscillation phenomenology will occur above a model-dependent energy threshold, and thus experiments capable of detecting extremely high energy neutrinos are particularly useful in such measurements.

The effects of quantum decoherence may also be observable in cosmic neutrinos. In particular, anti-neutrinos generated in the decays of neutrons produced in the photo-disintegration of ultra-high energy cosmic nuclei will have their flavor ratios significantly modified if quantum decoherence effects are significant. The very long baselines over which such neutrinos travel before reaching Earth provide an opportunity to test for these effects with much greater precision than other techniques can achieve.

\vspace{0.5cm}

{\bf Acknowledgements}: 

The work of DM is supported by a studentship from the University of
Sheffield.  The work of EW is supported by PPARC, grant reference number
PPA/G/S/2003/00082, the Royal Society and the London Mathematical Society.
EW would like to thank the Universities of Durham and Newcastle-upon-Tyne
and University College Dublin for hospitality while this work was completed. DH is supported by the Leverhulme trust.

\end{document}